\documentclass [10pt,a4paper]{article}
\usepackage[numbers,sort&compress]{natbib}
\usepackage{amsfonts}
\usepackage{indentfirst}
\usepackage{amsmath}
\usepackage{amsfonts}
\usepackage{amssymb}
\usepackage{amsthm}
\usepackage{maplestd2e}
\usepackage{a4wide}
\usepackage[CJKbookmarks]{hyperref}
\usepackage{graphicx}
\usepackage{float}
\usepackage{subfigure}

\def\Rmn#1{\expandafter\uppercase\expandafter{\romannumeral #1}}

\begin{document}
\newtheorem{theorem}{Theorem}[section]
\newtheorem{definition}{Def\mbox{}inition}[section]
\newtheorem{lemma}{Lemma}[section]
\newtheorem{corollary}{Corollary}[section]

\title{The weighted dif\mbox{}ference substitutions and Nonnegativity Decision of Forms\footnote{ Partially supported by a
National Key Basic Research Project of China (2004CB318000) and by
National Natural Science Foundation of China (10571095). Sponsored
by K.C.Wong Magna Fund in Ningbo University. }}
\date{}

\author{
 Xiaorong Hou$^1$~~ Song Xu$^{1,2}$~~Junwei Shao$^1$\\
 \textit{\small $1.$ College of Automation,
University of Electronic Science and Technology of China, Sichuan,
PRC}\\
 \textit{\small $2.$ Faculity of Science, Ningbo University, Ningbo,
Zhejiang, PRC}\\
\textit{\small E-mail:
\href{mailto:houxr@uestc.edu.cn}{houxr@uestc.edu.cn},
\href{mailto:xusong@nbu.edu.cn}{xusong@nbu.edu.cn} }}

 \maketitle
{\noindent\small  {\bf Abstract} In this paper, we study the
weighted dif\mbox{}ference substitutions from geometrical views.
First, we give the geometric meanings of the weighted
dif\mbox{}ference substitutions, and introduce the concept of
convergence of the sequence of substitution sets.    Then it is
proven that the
 sequence of the successive
weighted dif\mbox{}ference substitution sets is convergent.  Based
on the convergence
 of the sequence of the successive weighted
dif\mbox{}ference sets,  a new, simpler method to prove that if the
form  $F$ is positive def\mbox{}inite on $\mathbb{T}_n$, then the
sequence of  sets $\{\textrm{SDS}^{(m)}(F)\}_{m=1}^\infty$ is
positively terminating is presented,  which is dif\mbox{}ferent from
the one given in \cite{Yong:1}. That is, we can decide the
nonnegativity of a positive definite form by
 successively running the weighted
dif\mbox{}ference substitutions finite times. Finally, an algorithm
for deciding an indef\mbox{}inite form with a counter-example  is
obtained, and
 some examples are listed  by using the obtained algorithm.  }\\[2ex]

\\[2ex]


\section{ Introduction }
Theories and methods of nonnegative polynomials have been  widely
used in  robust control, non-linear control and non-convex
optimization \cite{P:1,P:2,J:1}, etc.   Some famous research works
on nonnegativity decision of polynomials without cell-decomposition
were given  by P\'{o}lya's Theorem \cite{Po:1,G:1} and papers
\cite{Ca:1,H:2}.

Few years ago,  Yang \cite{Yang:1,Yang:2,Yang:3} introduced a
heuristic method for nonnegativity  decision of polynomials, which
is now called Successive Dif\mbox{}ference Substitution (SDS). It
has been applied to prove a great many polynomial inequalities with
more variables and higher degrees. Yang recommended  further studies
on
 SDS and  put forward some open problems.

 A valuable progress on the
 topic was made by Yao \cite{Yong:1}.  He investigated a weighted dif\mbox{}ference
 substitution
 \begin{equation} \label{eq:12}
\left\{ \begin{aligned}
         x_1 &= t_1+\frac{t_2}{2}+\cdots+\frac{t_n}{n}, \\
         x_2 &= \frac{t_2}{2}+\cdots+\frac{t_n}{n},\\
             & \cdots\cdots\\
         x_n &= \frac{t_n}{n},
                          \end{aligned} \right.
                         \end{equation}
instead of the  original dif\mbox{}ference
 substitution

 \begin{equation} \label{eq:222}
\left\{ \begin{aligned}
         x_1 &= t_1+t_2+\cdots+t_n, \\
         x_2 &= t_2+\cdots+t_n,\\
             & \cdots\cdots\\
         x_n &= t_n,
                          \end{aligned} \right.
                         \end{equation}
and proved that, for a form (namely, a homogeneous polynomial) which
is positive definite on $\mathbb{R}_+^n$, the corresponding sequence
of SDS sets is positively terminating, where
$\mathbb{R}_+^n=\{(x_1,x_2,...,x_n)|x_i\geq 0, i=1,2,...,n\}.$  That
is,  we can decide the nonnegativity of a positive definite form by
 successively running SDS finite times.

 This paper is organized as follows. Section 2
introduces some preliminary notions of the weighted
dif\mbox{}ference substitutions. Section 3 provides a new
perspective  to study the weighted dif\mbox{}ference substitutions,
gives the geometric meanings  of them, and  proves that the sequence
of the successive  weighted dif\mbox{}ference
 substitution sets
is convergent.  A new, simpler method to prove that if the form  $F$
is positive def\mbox{}inite on $\mathbb{R}_+^n$, then the sequence
of sets $\{\textrm{SDS}^{(m)}(F)\}_{m=1}^\infty$ is positively
terminating is presented in Section 4, and an algorithm for deciding
an indef\mbox{}inite form with a counter-example in Section 5. By
using the obtained algorithm, several examples are listed in Section
6.


\section{Preliminary  notions}

We first introduce some notations and def\mbox{}initions according
to \cite{Yong:1} (with some differences).

  Consider $W_n\in \mathbb{R}^{n\times n}$,  where
\begin{equation}\label{eee}
W_n = \left[
\begin{array}{cccc}
1 & \frac{1}{2}& \ldots & \frac{1}{n} \\
0 & \frac{1}{2} & \ldots & \frac{1}{n}\\
\vdots & \ddots& \ddots& \vdots\\
0 & \ldots & 0 & \frac{1}{n}
\end{array} \right].
\end{equation}

Let $[k_1k_2\cdots k_n]$ be a permutation of $1, 2, \cdots, n$.
$P_{[k_1k_2\cdots k_n]}= [a(i,j)]$ is an $n \times n$ matrix for
which $a(1,k_1) = 1, a(2,k_2) = 1,\cdots, a(n,k_n) = 1, $ and 0 in
all other positions (Permutation matrix).

\begin{definition}
\emph{ $n \times n$ square matrix $B_{[k_1k_2\cdots k_n]}$ is
defined as follows:
 \begin{displaymath}
B_{[k_1k_2...k_n]}=P_{[k_1k_2...k_n]}W_n.
\end{displaymath}
And the set that consists of all $B_{[k_1k_2...k_n]}$ is denoted by
$PW_n$ (there are $n!$ elements in $PW_n$) and called the weighted
dif\mbox{}ference
 substitution matrix set.   Accordingly, the set of
linear transformations
\begin{displaymath}
\{X^{\textmd{Tr}}=B_{[\alpha]}T^{\textmd{Tr}}|B_{[\alpha]}\in
PW_n\},
\end{displaymath} is called  the weighted dif\mbox{}ference  substitution
set, which consists of $n!$ substitutions, where $X, T\in
\mathbb{R}_{+}^{n}$, and $X^{\textmd{Tr}}$, $T^{\textmd{Tr}}$ are
respectively the transposes of $X$, $T$. }
 \end{definition}

 \begin{definition}\label{zcyydh}
\emph{ The
 set of
linear transformations
\begin{displaymath}
\{X^{\textmd{Tr}}=B_{[\alpha_1]}B_{[\alpha_2]}\cdots
B_{[\alpha_m]}T^{\textmd{Tr}}|B_{[\alpha_i]}\in PW_n\},
\end{displaymath}
is called the $m$-times successive weighted dif\mbox{}ference
substitution set, which consists of $(n!)^m$  substitutions.}
\end{definition}

\begin{definition}\emph{Given the form $F\in R[x_1, x_2, \dots, x_n]$, when $[a_1], [a_2],\cdots,[a_m]$ traverse all the permutations of $1,
2,\cdots,n$ respectively, we def\mbox{}ine the set
\begin{displaymath}
\textrm{SDS}^{(m)}(F)=\bigcup\limits_{[\alpha_m]}^{n!}\cdots
\bigcup\limits_{[\alpha_{2}]}^{n!}\bigcup\limits_{[\alpha_1]}^{n!}F(B_{[\alpha_1]}B_{[\alpha_2]}\cdots
B_{[\alpha_m]}X^{\textmd{Tr}}),
\end{displaymath}
which is called  the $m$-times successive weighted dif\mbox{}ference
substitution set of the form $F$.}
 \end{definition}

\begin{definition}
\emph{  We def\mbox{}ine the sequence of sets
$\{\textrm{SDS}^{(m)}(F)\}_{m=1}^\infty$ as follows}
\begin{displaymath}
\{ \emph{\textrm{SDS}}(F)^{(m)}\}_{m=1}^\infty =
\emph{\textrm{SDS}}(F), \emph{\textrm{SDS}}^{(2)}(F), \cdots .
\end{displaymath}
 \end{definition}

It's time to def\mbox{}ine the termination of
$\{\textrm{SDS}^{(m)}(F)\}_{m=1}^\infty$, which is directly related
to the positive semi-definite property of the form $F$.

Let $\alpha=(\alpha_1,\alpha_2,\cdots,\alpha_n)\in\mathbb{N}^{n}$,
and let $|\alpha|=\alpha_1+\alpha_2+\cdots \alpha_n$. Then we write
a form $F$ with degree $d$ as $$F=\sum\limits_{|\alpha|=d}c_\alpha
x_1^{\alpha_1}x_2^{\alpha_2}\cdots x_n^{\alpha_n}.$$

 \begin{definition}
\emph{ The form $F$ is called trivially positive if the
coef\mbox{}f\mbox{}icients $c_\alpha$ of every term
$x_1^{\alpha_1}x_2^{\alpha_2}\cdots x_n^{\alpha_n}$ in $F$  are
nonnegative.  If $F(1,1,\cdots,1)<0$, then $F$ is called  trivially
negative.}
\end{definition}

\begin{definition}
   \emph{ If the form
   $F(X)\geq0$  for all $X\in\mathbb{R}_+^n$,  then $F$ is called positive semi-def\mbox{}inite ; If
   $F(X)>0$  for all $X(\neq0)\in\mathbb{R}_+^n$,
   $F$ is called positive def\mbox{}inite on
   $\mathbb{R}_+^n$; If there are $X$ and $Y\in \mathbb{R}_+^n$
such that $F(X)>0$ and  $F(Y)<0$,
   $F$ is called indef\mbox{}inite on
   $\mathbb{R}_+^n$. }
\end{definition}

  \begin{lemma}  \label{zpft}
\emph{Given the form $F$ on   $\mathbb{R}_+^n$, if the form $F$ is
trivially positive, then  $F$ is positive semi-def\mbox{}inite on
   $\mathbb{R}_+^n$; If the
form $F$ is trivially negative, then  $F$ isn't positive
semi-def\mbox{}inite on
   $\mathbb{R}_+^n$. }
\end{lemma}

\begin{definition}\label{deft}
\emph{Given a form $F$ on   $\mathbb{R}_+^n$, if there is a positive
integer $k$ such that every element of the set
$\textmd{SDS}^{(k)}(F)$ is trivially positive, the sequence of sets
$\{\textrm{SDS}^{(m)}(F)\}_{m=1}^\infty$ is called positively
terminating;  If there is a positive integer $k$ and a form $G$ such
that $G\in \textmd{SDS}^{(k)}(F)$ and $G$ is trivially negative, the
sequence of sets $\{\textrm{SDS}^{(m)}(F)\}_{m=1}^\infty$ is called
negatively terminating;   The sequence of sets
$\{\textrm{SDS}^{(m)}(F)\}_{m=1}^\infty$ is neither positively
terminating nor negatively terminating, then it is called not
terminating.}
 \end{definition}

  By Definition 2.7, it's easy to get the following lemma.

    \begin{lemma}
\emph{Given the form $F$ on   $\mathbb{R}_+^n$, if the sequence of
sets $\{\textrm{SDS}^{(m)}(F)\}_{m=1}^\infty$ is positively
terminating, then  $F$ is positive semi-def\mbox{}inite on
   $\mathbb{R}_+^n$; If the
sequence of sets $\{\textrm{SDS}^{(m)}(F)\}_{m=1}^\infty$ is
negatively terminating, then  $F$ isn't positive
semi-def\mbox{}inite on
   $\mathbb{R}_+^n$. }
\end{lemma}

Next, we will give the def\mbox{}inition of the normalized
substitution.

\begin{definition}
   \emph{Let $V=[v_{ij}]$ be an $n\times n$ matrix. If $\sum\limits_{i=1}^nv_{ij}=1, j=1,2,\cdots
   n$, $V$ is called a normalized matrix.  And the corresponding substitution
   \begin{displaymath}
X^{\textmd{Tr}}=VT^{\textmd{Tr}},
\end{displaymath}}
\emph{is called a normalized substitution.}
    \end{definition}

\begin{lemma}  \label{zc}
\emph{Let $ U = [u_{ij}] = V_1 V_2 \cdots V_k$. If
$V_1,V_2,\cdots,V_k$ are normalized matrices, then $U$ is a
normalized matrix , that is,
 \begin{displaymath}
 \sum\limits_{i=1}^nu_{ij}=1(j=1,2,\cdots n).
 \end{displaymath}
 And let
 $(x_1,x_2,\cdots,x_n)^{\textmd{Tr}}=U(t_1,t_2,\cdots,t_n)^{\textmd{Tr}}$,
 then
 $\sum\limits_{i=1}^{n}x_i=1$ iff
 $\sum\limits_{i=1}^{n}t_i=1$.}
\end{lemma}
 The proof of Lemma  \ref{zc} is very straightforward and is omitted.

The $(n-1)$-dimensional simplex  is defined as follows
$$\mathbb{T}_n = \{(x_1,x_2,\cdots,x_n)|\sum\limits_{i=1}^nx_i=1 , (x_1,x_2,\cdots,x_n)\in\mathbb{R}_{+}^{n}\}.$$

\begin{definition}
   \emph{ If the form
   $F(X)\geq0$  for all $X\in\mathbb{T}_n$,  then $F$ is called positive semi-definite on
   $\mathbb{T}_n$;
    If
   $F(X)>0$  for all $X\in\mathbb{T}_n$, then
   $F$ is called positive definite on
   $\mathbb{T}_n$; If there are $X$ and $Y\in \mathbb{T}_n$
such that $F(X)>0$ and  $F(Y)<0$,  then
   $F$ is called indefinite on
   $\mathbb{T}_n$. }
\end{definition}

     Obviously, we have the following conclusion.
 \begin{lemma}  \label{abcd}
\emph{The form $F$ is positive semi-definite (positive definite,
indefinite) on
   $\mathbb{T}_n$ iff $F$ is positive semi-definite (positive definite,
indefinite) on
   $\mathbb{R}_+^n$.}
\end{lemma}

According to Lemma \ref{abcd}, for brevity, we suppose that the form
$F$ is defined  on $\mathbb{T}_n$ in the remainder of this paper.

\section{ Convergence
 of the sequence of successive weighted dif\mbox{}ference substitution sets.}
 In this section, we'll consider the  weighted dif\mbox{}ference
substitutions from geometrical
 views.

 Let $X=(x_1,x_2,\cdots,x_n)\in
\mathbb{T}_n$. Consider the weighted dif\mbox{}ference substitution
  \begin{equation}\label{yangyaos}
X^{\textmd{Tr}}=W_nT^{\textmd{Tr}} ,
  \end{equation}
where $W_n$ is denoted by (\ref{eee}).

 By (\ref{yangyaos}), if
$X=A_1=(1,0,...,0)$, then $T=(1,0,...,0)$; If
$X=A_2=(\frac{1}{2},\frac{1}{2},...,0)$, then $T=(0,1,...,0)$;
$\cdots$; If $X=A_n=(\frac{1}{n},\frac{1}{n},...,\frac{1}{n})$, then
$T=(0,0,...,1)$. Moreover, $A_k $ is the barycenter of the
(k-1)-dimentional proper face of $\mathbb{T}_n$ which contains the
points $A_1, A_2, \cdots, A_k$ for $k=1,2,\cdots,n$ .  According to
Lemma \ref{zc}, for all $(x_1,x_2,\cdots,x_n)\in \mathbb{T}_n$, we
have $t_1+t_2+\cdots +t_n=1$. Therefore, $A_1A_2\cdots A_n$ is a
subsimplex of the first barycentric subdivision of $ \mathbb{T}_n$,
satisfying
$W_n=[A_1^{\textmd{Tr}},A_2^{\textrm{Tr}},\cdots,A_n^{\textmd{Tr}}]$.
And it is indicated that the weighted dif\mbox{}ference substitution
(\ref{yangyaos}) and  the subsimplex $A_1A_2\cdots A_n$ correspond
to each other.

 Analogously,  the other $n!-1$ weighted dif\mbox{}ference substitutions correspond
 to  the other
 $n!-1$ subsimplexes of the first barycentric subdivision of $\mathbb{T}_n$.

  Hence, from geometrical
 views, the weighted dif\mbox{}ference substitution set corresponds to
  the f\mbox{}irst barycentric subdivision of $\mathbb{T}_n$, that is,  there is a one-to-one correspondence between the weighted dif\mbox{}ference substitutions and the  subsimplexes of the first
barycentric subdivision of $ \mathbb{T}_n$.
 \begin{figure}[H]
\begin{center}
\includegraphics[width=0.25\textwidth]{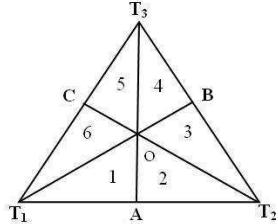}
 \caption{Barycentric subdivision}\label{Fig1}
\end{center}
\end{figure}

  For instance, when $n=3$, the following six weighted dif\mbox{}ference substitution
  matrices correspond to subsimplexes labeled as 1-6 in Fig. 1,
  respectively,

\begin{displaymath}
\left[
\begin{array}{lll}
1 & \frac{1}{2} & \frac{1}{3} \\
0 & \frac{1}{2} & \frac{1}{3} \\
0 & 0 & \frac{1}{3}
\end{array} \right],
 \left[
\begin{array}{lll}
0 & \frac{1}{2} & \frac{1}{3} \\
1& \frac{1}{2} & \frac{1}{3} \\
0 & 0 & \frac{1}{3}
\end{array} \right],\left[
\begin{array}{lll}
0 & 0 & \frac{1}{3} \\
1 & \frac{1}{2} & \frac{1}{3} \\
0 & \frac{1}{2} & \frac{1}{3}
\end{array} \right],
\left[
\begin{array}{lll}
0 & 0 & \frac{1}{3} \\
0 & \frac{1}{2} & \frac{1}{3} \\
1 & \frac{1}{2} & \frac{1}{3}
\end{array} \right],
\left[
\begin{array}{lll}
0 & \frac{1}{2} & \frac{1}{3} \\
0 &  0 & \frac{1}{3} \\
1 & \frac{1}{2} & \frac{1}{3}
\end{array} \right],
\left[
\begin{array}{lll}
1 & \frac{1}{2} & \frac{1}{3} \\
0 & 0 & \frac{1}{3} \\
0 & \frac{1}{2} & \frac{1}{3}
\end{array} \right].
\end{displaymath}

By Lemma \ref{zc} and Def\mbox{}inition \ref{zcyydh}, we have that
the $m$-times successive weighted dif\mbox{}ference  substitution
set corresponds to the $m$-th barycentric subdivision of
$\mathbb{T}_n$.

  Next, we'll introduce the concept of
convergence of the sequence of  substitution sets.

\begin{definition}\label{sld1} \emph{ Let $\sigma$ be a subsimplex of $\mathbb{T}_n$,  the maximum distance between
 vertexs of $\sigma$
is called the diameter of $\sigma$.}
\end{definition}

\begin{definition}\label{sld0}
\emph{ Let $K_0=\mathbb{T}_n$, and $K_{i+1}$ be the subdivision of
$K_{i}$ for $i=1,2,\cdots.$  If for all $ \varepsilon>0$, there
exists $N\in \mathbb{N}$ such that all the diameters of the
subsimplexes
 of $K_N$ are less than $\varepsilon$, the subdivision sequence
$\{K_i\}_{i=1}^\infty$ is called convergent.}
\end{definition}

\begin{definition}\label{sld5}
\emph{Suppose that the subdivision scheme through which $K_{i-1}$ is
subdivided into $K_{i}$ corresponds to the substitution set $L_{i}$
for
 $i=1,2,\cdots$. If the sequence $\{K_i\}_{i=1}^\infty$ is convergent, the
 sequence of substitution sets
$\{L_i\}_{i=1}^\infty$ is called convergent. If $L_{i}=L$,
$i=1,2,\cdots$, brief\mbox{}ly, we say the sequence of the
successive $L$-substitution sets is convergent.}
\end{definition}
It's time to consider  the convergence of the sequence of the
successive weighted dif\mbox{}ference substitution sets.
\begin{lemma}\label{bs}
\emph{\cite{Edwin:1, James:1} Let $K$ be a complex. If $K_N$ is the
$k$-th barycentric subdivision of $K$, then for all $\varepsilon>0$,
there exists $N\in \mathbb{N}$ such that all the diameters of the
subsimplexes
 of $K_N$ are less than $\varepsilon$. }
\end{lemma}

 By Lemma \ref{bs} , we have the following theorem,  which plays important
 roles
 in the proofs of Theorem \ref{zd} and Theorem \ref{ce} in the next sections.

\begin{theorem}
\emph{The   barycentric subdivision
 sequence $\{K_m\}_{m=1}^\infty$ of $ \mathbb{T}_n$ and the corresponding sequence of
the successive weighted dif\mbox{}ference substitution sets
$\{\textrm{SDS}^{(m)}(F)\}_{m=1}^\infty$
 are convergent.} \label{sl}
\end{theorem}

\section{Nonnegativity decision of forms }
Given a form $F$ on $\mathbb{T}_n$. We know that if the sequence of
sets $\{\textrm{SDS}^{(m)}(F)\}_{m=1}^\infty$ is negatively
terminating, then we can conclude that $F$  is positive
semi-definite on $\mathbb{T}_n$. Thus there is a natural question,
that is, which kind of forms can be solved by the method? Yao
\cite{Yong:1} proved that, for a positive definite form $F$ on
$\mathbb{T}_n$, the sequence of sets
$\{\textrm{SDS}^{(m)}(F)\}_{m=1}^\infty$ is positively terminating.

In this section, we present a new, simpler method to prove the
conclusion, which is based on the convergence
 of the sequence of the successive weighted dif\mbox{}ference substitution sets.

\begin{theorem} \label{zd}
\emph{Let the form  $F$ be positive def\mbox{}inite on
$\mathbb{T}_n$, then the sequence of  sets
$\{\textrm{SDS}^{(m)}(F)\}_{m=1}^\infty$ is positively terminating.}
\end{theorem}
\begin{proof} We only give the proof for the ternary form with
degree $d$, and the multivariate form can be gotten by induction.

Suppose that
$F(x_1,x_2,x_3)=\sum\limits_{i+j+k=d}a_{ijk}x_1^ix_2^jx_3^k$. An
arbitrary $m$-times successive weighted dif\mbox{}ference
substitution can be written as
\begin{equation} \label{eq11}
\left\{ \begin{array}{l} x_1  =  k_{1}t_1+(k_{1}+\alpha_1)t_2+(k_{1}+\beta_1)t_3,\\
x_2  =  k_{2}t_1+(k_{2}+\alpha_2)t_2+(k_{2}+\beta_2)t_3,\\
x_3  =  k_{3}t_1+(k_{3}+\alpha_3)t_2+(k_{3}+\beta_3)t_3,
\end{array}\right.
\end{equation}
where $ \sum\limits_{i=1}^3k_i =1, \sum\limits_{i=1}^3\alpha_i = 0$
and $ \sum\limits_{i=1}^3\beta_i=0 $.

Let $t=t_1+t_2+t_3$, then (\ref{eq11})  becomes
\begin{equation}
\left\{ \begin{array}{l} x_1  = k_{1}t+\alpha_1t_2+\beta_1t_3,\\
x_2  =  k_{2}t+\alpha_2t_2+\beta_2t_3,\\
x_3  =  k_{3}t+\alpha_3t_2+\beta_3t_3.
\end{array}\right.
\end{equation}
Thus
\begin{equation}
\begin{split}
\Phi(t_1,t_2,t_3) & = F(k_{1}t+\alpha_1t_2+\beta_1t_3,
k_{2}t+\alpha_2t_2+\beta_2t_3,k_{3}t+\alpha_3t_2+\beta_3t_3)\\
 & =
\sum\limits_{i+j+k=d}a_{ijk}(k_{1}t+\alpha_1t_2+\beta_1t_3)^i(k_{2}t+\alpha_2t_2+\beta_2t_3)^j(k_{3}t+\alpha_3t_2+\beta_3t_3)^k \\
&=
\sum\limits_{i+j+k=d}a_{ijk}(k_{1}^it^i+\sum\limits_{p+q+r=i,p\neq
i}\frac{i!}{p!q!r!}k_1^p\alpha_1^q\beta_1^rt^pt_2^qt_3^r)\cdot\\
&(k_{2}^jt^j+\sum\limits_{p+q+r=j,p\neq
j}\frac{j!}{p!q!r!}k_2^p\alpha_2^q\beta_2^rt^pt_2^qt_3^r)(k_{3}^it^i+\sum\limits_{p+q+r=k,p\neq
k}\frac{k!}{p!q!r!}k_3^p\alpha_3^q\beta_3^rt^pt_2^qt_3^r)
\\
&
=\sum\limits_{i+j+k=d}a_{ijk}k_{1}^ik_{2}^jk_{3}^kt^d+\sum\limits_{i+j+k=d}\phi_{ijk}(k_{1},k_{2},k_{3},\alpha_1,\alpha_2,\alpha_3,\beta_1,\beta_2,\beta_3)t_1^it_2^jt_3^k\\
&
=F(k_{1},k_{2},k_{3})t^d+\sum\limits_{i+j+k=d}\phi_{ijk}(k_{1},k_{1},k_{3},\alpha_1,\alpha_2,\alpha_3,\beta_1,\beta_2,\beta_3)t_1^it_2^jt_3^k\\
&
=\sum\limits_{i+j+k=d}\left({\frac{d!}{i!j!k!}}F(k_{1},k_{2},k_{3})+\phi_{ijk}(k_{1},k_{2},k_{3},\alpha_1,\alpha_2,\alpha_3,\beta_1,\beta_2,\beta_3)\right)t_1^it_2^jt_3^k\\
& =\sum\limits_{i+j+k=d}A_{ijk}t_1^it_2^jt_3^k.
\end{split}
\end{equation}
where
\begin{equation}\label{aaa}
\begin{split}
A_{ijk}={\frac{d!}{i!j!k!}}F(k_{1},k_{2},k_{3})+\phi_{ijk}(k_{1},k_{2},k_{3},\alpha_1,\alpha_2,\alpha_3,\beta_1,\beta_2,\beta_3).
\end{split}
\end{equation}

Obviously,
\begin{equation}\label{bbb}
\begin{split}
\lim_{(\alpha_1,\alpha_2,\alpha_3,\beta_1,\beta_2,\beta_3)\rightarrow(0,0,0,0,0,0)}\phi_{ijk}(k_{1},k_{2},k_{3},\alpha_1,\alpha_2,\alpha_3,\beta_1,\beta_2,\beta_3)=0.
\end{split}
\end{equation}
And  since  $F(x_1,x_2,x_3)$ is positive def\mbox{}inite on
$\mathbb{T}_n$,  there exists $\varepsilon>0$ such that
\begin{equation}\label{ccc}
F(k_{1},k_{2},k_{3})\geq\varepsilon>0. \end{equation}

 On the one hand,   the  vertexs of the subsimplex which corresponds to the  successive weighted dif\mbox{}ference
 substitution (\ref{eq11})  are respectively
 $$(k_1,k_2,k_3),
(k_1+\alpha_1,k_2+\alpha_2,k_3+\alpha_3),
(k_1+\beta_1,k_2+\beta_2,k_3+\beta_3).$$
By Theorem (\ref{sl}),
 $\alpha_1,\alpha_2,\alpha_3,\beta_1,\beta_2,\beta_3$ can be suf\mbox{}f\mbox{}iciently small
 when $m$ is  suf\mbox{}f\mbox{}iciently  large.

 On the other hand, for $F$ is continuous on $\mathbb{T}_n$ and by
 (\ref{aaa})-(\ref{ccc}), we have $A_{ijk}>
 0$ when $\alpha_1,\alpha_2,\alpha_3,\beta_1,\beta_2,\beta_3$ are suf\mbox{}f\mbox{}iciently
 small.

Putting together the above two aspects,  we have that  there exists
a  suf\mbox{}f\mbox{}iciently  large integer $m$  such that $F$
becomes trivially positive by (\ref{eq11}).
 For the successive weighted dif\mbox{}ference substitution (\ref{eq11}) is arbitrary, the sequence of  sets $\{\textrm{SDS}^{(m)}(F)\}_{m=1}^\infty$ is
positively terminating. \end{proof}

According to the proof of Theorem  \ref{zd}, we obtain the following
conclusion.

\begin{corollary}\label{zd111}
\emph{Let the form $F$ be positive definite on $\mathbb{T}_n$,  then
by an arbitrary $m$-times successive weighted dif\mbox{}ference
substitution,  when $m$ is suf\mbox{}f\mbox{}iciently larger,  $F$
can become a nonlacunary  trivially positive  form.} \label{gzd}
\end{corollary}

Theorem  \ref{zd} is somewhat analogous to  P\'{o}lya's Theorem.
However,  many examples show that P\'{o}lya's Theorem seems almost
useless to positive semi-definite forms except for few cases, while
SDS is demonstrated very helpful to positive semi-definite ones as
well.

\section{Decision of indef\mbox{}inite forms}

Many problems,  such  as the inequality disproving, are always
transformed into decision of indef\mbox{}inite forms.

 Given a form
$F$ on $\mathbb{T}_n$. Suppose that there exists $X_0\in
\mathbb{T}_n$, such that $F(X_0)>0$. It is well-known to us that if
the sequence of sets $\{\textrm{SDS}^{(m)}(F)\}_{m=1}^\infty$ is
negatively terminating, then $F$  is indef\mbox{}inite on
$\mathbb{T}_n$. Then it follows a question naturally: for an
indef\mbox{}inite form $F$ on $\mathbb{T}_n$, is the
$\{\textrm{SDS}^{(m)}(F)\}_{m=1}^\infty$ negatively terminating? The
following theorem answers the question.

\begin{theorem} \label{ce}
\emph{Let the form $F$ be indef\mbox{}inite on $\mathbb{T}_n$,  then
the sequence of sets $\{\textrm{SDS}^{(m)}(F)\}_{m=1}^\infty$ is
negatively terminating.}
\end{theorem}

\begin{proof}  Since the form $F$ is indef\mbox{}inite on
   $\mathbb{T}_n$,  there exists $X_0\in
\mathbb{T}_n$ such that $F(X_0)<0$ .  And $F$ is continuous on
$\mathbb{T}_n$,  so there exists a neighborhood
$\textrm{U}(X_0)\subset \mathbb{T}_n$ of $X_0$ (If $X_0$ is on the
boundary of $\mathbb{T}_n$, then we take $\textrm{U}(X_0)\bigcap
\mathbb{T}_n$) such that  $F(X)<0$ for all $X\in \textrm{U}(X_0)$.
For the  barycentric subdivision
 sequence of $ \mathbb{T}_n$ is
convergent, then there exists a subsimplex $\sigma$  of the $k$-th
barycentric subdivision of $\mathbb{T}_n$, which corresponds to the
$k$-times successive weighted dif\mbox{}ference substitution

 $$X^{\textmd{Tr}}=B_{[i_1]}B_{[i_2]}\cdots B_{[i_k]}T^{\textmd{Tr}},~~ B_{[i_1]}, B_{[i_2]}, \cdots,B_{[i_k]}\in PW_n,$$
 satisfying $\sigma
\subset\textrm{U}(X_0)$, where $k$ is a  suf\mbox{}f\mbox{}iciently
larger integer, and $PW_n$ is the weighted dif\mbox{}ference
 substitution matrix set.  Thus, $-F(X)$ is positive definite on
 $\sigma$. By Theorem  \ref{zd},
the sequence of  sets $\{\textrm{SDS}^{(m)}(-F)\}_{m=1}^\infty$ is
positively terminating, so there exists an
 $l$-times successive
weighted dif\mbox{}ference substitution
 $$X^{\textmd{Tr}}=B_{[j_1]}B_{[j_2]}\cdots
B_{[j_l]}T^{\textmd{Tr}},  ~~ B_{[j_1]}, B_{[j_2]},
\cdots,B_{[j_l]}\in PW_n,$$
 satisfying that $F(B_{[i_1]}B_{[i_2]}\cdots B_{[i_k]}B_{[j_l]}B_{[j_2]}\cdots B_{[j_l]}T^{\textmd{Tr}})$
 is trivially negative. Therefore, the sequence of sets $\{\textrm{SDS}^{(m)}(F)\}_{m=1}^\infty$ is
negatively terminating.\end{proof}


By the proving process of Theorem \ref{ce},  we obtain the following
algorithm,  which is used to decide the  nonnegativity of a form or
to decide an indef\mbox{}inite form with a
counter-example.\\
\textbf{Algorithm} \textbf{(YYS)}\\
Input: the form  $F\in \mathbb{Q}[x_1,x_2,\cdots, x_n]$, where  $F$ is positive def\mbox{}inite or indef\mbox{}inite on $ \mathbb{R}_{+}^n$ .  \\
Output:  ``The form $F$ is positive semi-def\mbox{}inite'', or ``$\tilde{X}_0$, $F(\tilde{X}_0)<0$''.\\
step1: Let  $\mathbb{F}=\{F\}$.\\
step2:  Compute $\bigcup\limits_{F\in\mathbb{F}}\textrm{SDS}(F)$,

\qquad Let
$$\mathbb{F}=\bigcup\limits_{F\in\mathbb{F}}\textrm{SDS}(F)-\{\textmd{trivially
positive forms in}
\bigcup\limits_{F\in\mathbb{F}}\textrm{SDS}(F)\}\triangleq\{F_{[1]},F_{[2]},\cdots,F_{[k]}\},$$

\qquad where
$$F_{[i]}= F(B_{[i]}X^{\textmd{Tr}}),\quad B_{[i]} \in PW_n.$$
 step3: Let $L=[[1],[2],\cdots,[k]]$.

\quad step31:  If $\mathbb{F}$ is null, then output ``the form $F$
is positive semi-definite'', and terminate.

\quad step32:  If there is a trivially negative form
$F_{[i]}\in\mathbb{F}$, then output
$$ ``\tilde{X}_0 =
 B_{L[i]}(\frac{1}{n},\frac{1}{n},...,\frac{1}{n})^{\textmd{Tr}},~~F(\tilde{X}_0)<0\mbox{''}, $$

 ~~~~~~~~~~~~  and terminate, where

 $$B_{L[i]}=B_{[L[i][1]]}B_{[L[i][2]]}\cdots
B_{[L[i][m]]},$$  \quad \quad \quad \quad\qquad

 ~~~~~~~~~~~~  $L[i]$ is the $i$-th component of $L$, $L[i][j]$ is the $j$-th component
 of $L[i]$, and $m$ is the

  ~~~~~~~~~~~~  the total number of the components of $L[i]$.

\quad step33:  Else, Compute
$\bigcup\limits_{F\in\mathbb{F}}\textrm{SDS}(F)$.  Let

\begin{displaymath}
\begin{split}
\mathbb{F} &
=\bigcup\limits_{F\in\mathbb{F}}\textrm{SDS}(F)-\{\textmd{trivially
positive forms in}
\bigcup\limits_{F\in\mathbb{F}}\textrm{SDS}(F)\}\\
& \triangleq
\{F_{[\textrm{op}(L[1]),1]},\cdots,F_{[\textrm{op}(L[1]),l_1]},F_{[\textrm{op}(L[2]),1]},\cdots,F_{[\textrm{op}(L[2]),l_2]},\\
&
\cdots,F_{[\textrm{op}(L[k]),1]},\cdots,F_{[\textrm{op}(L[k]),l_k]}\},
\end{split}
\end{displaymath}

~~~~~~~~~~~~  where
$$F_{[\textrm{op}(L[i]),j]}=F(B_{[\textrm{op}(L[i]),j]}X^{\textmd{Tr}}),$$

~~~~~~~~~~~~   and $\textrm{op}(L[i])$ extracts operands from
$\textrm{op}(L[i])$. And let
\begin{displaymath}
\begin{split}
L& =[[\textrm{op}(L[1]),1],\cdots,[\textrm{op}(L[1]),l_1],[\textrm{op}(L[2]),1],\cdots,[\textrm{op}(L[2]),l_2],\\
& \cdots,[\textrm{op}(L[k]),1],\cdots,[\textrm{op}(L[k]),l_k]],
\end{split}
\end{displaymath}

~~~~~~~~~~~~ then go to step3.

By Algorithm YYS, we design  a Maple program called YYS, see
Appendix. To the program YYS,  there are some positive semi-definite
forms making the program do not terminate, that is, we cann't decide
these positive semi-definite forms by the method.

\section{Examples}

In this section, we demonstrate the program  YYS with some examples.

 \textbf{Example 1.} Show that the following form is positive semi-definite on
 $\mathbb{R}_{+}^3$,
 \begin{equation}
F(x,y,z)=x(x-y)^5-y(-z-y)^5-z(x-z)^5.
\end{equation}

Utilize the program YYS and execute order YYS($F$,[x,y,z]).  The
procedure need successively run the weighted dif\mbox{}ference
substitutions 3 times, then outputs:``The form $F$ is positive
semi-definite''.

\textbf{Example 2.} Show that the following form is
indef\mbox{}inite on $ \mathbb{R}_{+}^3$.
 \begin{equation}
\begin{split}
F(x,y,z)=7x^3-12x^2y-12x^2z+6xy^2+12xyz+6xz^2-\frac{9}{10}y^3-3y^2z-3yz^2-\frac{4}{5}z^3.
\end{split}
\end{equation}

   Executing order YYS($F$, [x,y,z]), we have a
counter-example by successively running the weighted
dif\mbox{}ference substitutions 2 times:
 $$\tilde{X}_0 =(
 \frac{37}{108}, \frac{49}{108},
 \frac{11}{54})^{\textmd{T}},
 \quad F(\tilde{X}_0)<0.$$
Obviously, $F(1,0,0)=7>0$, so the  form $F$ is indef\mbox{}inite on
$ \mathbb{R}_{+}^3$.

\textbf{Example 3.} Let $x\geq 0, y\geq 0, z\geq 0$, and $x+y+z\neq
0$. Try to decide whether the following inequality holds for all
$p\in\mathbb{N}$.
\begin{equation}\label{eq1}
\begin{split}
\frac{2}{3}(\frac{x^2}{y+z}+\frac{y^2}{z+x}+\frac{z^2}{x+y})-(\frac{x^p+y^p+z^p}{3})^{\frac{1}{p}}\geq
0.
\end{split}
\end{equation}

Take of\mbox{}f denominators of the left polynomial, and denote the
new polynomials by $F_1$,$F_2$,$\cdots$,$F_6$  for $p=1,
2,\cdots,6$, respectively. Execute order YYS($F_p$, [x,y,z]),
$p=1,2,...6$.  For $p=1, 2,\cdots,5$, the procedure only need run
the weighted dif\mbox{}ference substitutions 1 time, then outputs:
``The form $F$ is positive semi-definite''.  For $p=6$,  we have a
counter-example by successively running the weighted
dif\mbox{}ference substitutions 5 times:
 $$\tilde{X}_0 =(\frac{2159}{5832},
 \frac{3685}{11664},
 \frac{3661}{11664})^{\textmd{T}},
 \quad F_6(\tilde{X}_0)<0.$$ So
the  inequality cann't hold for $p=6$.

\subsection*{ Appendix.  Maple Program YYS}
\begin{maplettyout}
 YYS:=proc(poly,var)
 local a,b,A,f,i,j,k,m,n,r,s,t,F,G,H,M,W,newvar,st,Var:
 uses combinat, LinearAlgebra:
 F:=[[poly,[0]]]:
 n:=nops(var):
 Var:=convert(var,Vector):
 W:=(n)->Matrix(n,n,(i,j)->`if`(i<=j,1/j,0)):
 b:=permute(n): a:=W(n):
 A:=[seq(<seq(a[b[i][j]],j=1..n)>,i=1..n!)]:
 for i to nops(A) do
     for j to n do
         newvar[i,j]:=op(j,convert(A[i].Var,list)):
     od:
 od:
 r:=100:
 for s to r do
    m:=nops(F):
    f:=[]:
    for k to m do
        G[k]:=[]:
        for i from 1 to nops(A) do
            st:={seq(Var[j]=newvar[i,j],j=1..n)}:
            G[k]:=[op(G[k]),[expand(subs(st,F[k][1])),[op(F[k][2]),i]]]:
        od:
    od:
    F:=[seq(op(G[u]),u=1..nops(F))]:
    for i to nops(F) do
        if max([coeffs(F[i][1])]) < 0 then
            print(F[i][2]):
            M:=IdentityMatrix(n):
            for j from 2 to nops(F[i][2]) do
                M:=M.A[F[i][2][j]]:
            od:
            print(convert(M.Vector[column](n,1/n),list)):
            return ("The form is indefinite"):
       elif min([coeffs(F[i][1])]) > 0 then
            f:=[op(f),i]:
        fi:
    od:
    if nops(f)>0 then
        F:=subs({seq(F[f[t]]=NULL,t=1..nops(f))},F):
    fi:
    if nops(F)=0 then
        print(s):
        return("The form is positive semi-definite"):
    fi:
 od:
 end proc:

\end{maplettyout}

 \clearpage
\end{document}